\documentclass[graybox]{svmult}
\pdfoutput=1

\usepackage{type1cm}        
\usepackage{makeidx}         
\usepackage{graphicx}        
\usepackage{multicol}        
\usepackage[bottom]{footmisc}

\usepackage{newtxtext}       %
\usepackage{newtxmath}       

\usepackage{booktabs} 
\usepackage{arydshln} 
\usepackage{tabularx}
\usepackage{tabu}

\makeindex             

\begin{document}

\title*{Addressing the Recitative Problem \break in Real-time Opera Tracking}
\author{Charles Brazier and Gerhard Widmer}
\institute{Charles Brazier \at Institute of Computational Perception, Johannes Kepler University,\\
Altenberger Straße 69, 4040 Linz, Austria \\
\email{charles.brazier@jku.at}
\and Gerhard Widmer \at Institute of Computational Perception and LIT AI Lab, Linz Institute of Technology,\\
Altenberger Straße 69, 4040 Linz, Austria \\
\email{gerhard.widmer@jku.at}}
%
%
\maketitle

\abstract{Robust real-time opera tracking (score following) would be extremely useful for many processes surrounding live opera staging and streaming, including automatic lyrics displays, camera control, or live video cutting. Recent work has shown that, with some appropriate measures to account for common problems such as breaks and interruptions, spontaneous applause, various noises and interludes, current audio-to-audio alignment algorithms can be made to follow an entire opera from beginning to end, in a relatively robust way. However, they remain inaccurate when the textual content becomes prominent against the melody or music -- notably, during \textit{recitativo} passages. In this paper, we address this specific problem by proposing to use two specialized trackers in parallel, one focusing on music-, the other on speech-sensitive features. We first carry out a systematic study on speech-related features, targeting the precise alignment of corresponding recitatives from different performances of the same opera. Then we propose different solutions, based on pre-trained music and speech classifiers, to combine the two trackers in order to improve the global accuracy over the course of the entire opera.}

\section{Introduction}
\label{sec:introduction}

Score following aims at aligning a music performance to its corresponding score in real time, in order to retrieve the score position currently played by the musician(s) during a live performance. Since its start in the 1980s \cite{dannenberg1984line}, score following research has produced increasingly sophisticated music trackers, capable of reliably following complex solo performances and even full orchestras \cite{arzt2015real}, supporting such applications as automatic page turning for pianists \cite{arzt2008automatic}, live performance visualisation \cite{lartillot2020real}, or score viewing and automatic contextualization in orchestra concerts \cite{prockup2013orchestral,arzt2015artificial}.

Reliably tracking complex musical stage works such as full operas, however, is an even more difficult task. Operas are highly complex and heterogeneous works, involving not only a full orchestra superimposed with a complex mixture of singing voices (of highly expressive and varying singing or speaking style, generally with strong vibrato), but also various acting breaks and interruptions, spontaneous applause, stage noises, interludes, etc. At the same time, robust live opera tracking systems would be extremely useful for many processes surrounding live opera staging and streaming, including automatic lyrics displays, camera control, or live video cutting.

A first step towards robust tracking of full operas was presented in \cite{brazier2020towards}. The approach to score following taken there makes use of an audio-to-audio alignment strategy and aligns a live opera, the \textit{target}, with a \textit{reference recording}, which was aligned to the printed score before and serves as a proxy to the score. This method has the advantage of aligning two audio signals that contain realistic sounds and that can be expected to be more similar, compared to using an artificial reference audio computed from a symbolic score (given as MIDI or Music XML). It also obviates the need for full symbolic opera scores, which are generally not available. However, it does require manual annotation beforehand. In \cite{brazier2020towards}, the real-time alignment between live target and reference is then computed with an adapted version of the \textit{On-Line Dynamic Time Warping (OLTW)} algorithm \cite{dixon2005line}, combined with applause, speech, and music classifiers that exert top-down control on the tracker in specific situations.

However, some open problems remain, notably tracking \textit{recitativo} sections, that are often introduced between songs to progress fast in the story. Recitatives are strongly acted and are dominated by the singers' voices. The singers are relatively free to impose their own rhythm to best match their acting on stage. The movements on stage alter directly the audio quality that is recorded by fixed microphones. Also, there may be high variability in vocal style, with singers singing, speaking, whispering, or laughing. In addition, the accompaniment is usually underspecified in the score and filled in / improvised in different ways (and even on different instruments) in different recordings.

In this paper, we will, after presenting the real-world musical material that we will be working with (Section \ref{sec:data}), present empirical results demonstrating the difficulty of recitative tracking (Section \ref{sec:baseline}), carry out a systematic search for audio features that are more effective for tracking recitatives (Section \ref{sec:feature_study}), and investigate several strategies for combining music- and speech-sensitive based trackers, using pre-trained audio classifiers (Section \ref{sec:combination}). The investigations are based on several full-length recordings of W.A.Mozart's opera \textit{Don Giovanni}, recorded live in an opera hall.

\section{Data description}
\label{sec:data}

As opera research datasets including recordings and corresponding annotations are not yet freely available, we had to compile our own dataset. As our first object of study, we chose the opera \textit{``Don Giovanni"} by W.A.Mozart (considered by some to be the ``opera of all operas''). We collected and manually annotated three complete recordings of this work, one from a commercial CD (to serve as the proxy to the score, the \textit{reference performance}), the other two recent live recordings from the opera hall, provided to us by the \textit{Vienna State Opera}\footnote{https://www.wiener-staatsoper.at/en/}. The latter two will be the target performances that we wish to align/track on-line. There are a few pieces (arias etc.) that are present in the reference but not in the target performances; these will be ignored for this study. The two target performances have the same structure.

\begin{table*}[t]
 \begin{center}
  \begin{tabu} to \textwidth {@{}l l *2{X[c]} c X[l] @{}}
  \toprule
  \textbf{Conductor} & \textbf{Orchestra} & \textbf{Year} & \textbf{Duration} & \textbf{Dur.~Recitatives} & \textbf{Role}\\
  \midrule
  H.v. Karajan & Berliner Philharmoniker & 1985 & 2:57:53 & 0:30:03 & Reference \\
  \'A. Fischer & Orchester der Wiener Staatsoper & 2018 & 3:12:54 & 0:34:58 & Target \\
  A. Manacorda & Orchester der Wiener Staatsoper & 2019 & 3:07:09 & 0:30:40 & Target \\
  \bottomrule
 \end{tabu}
\end{center}
 \caption{Dataset used in the audio-to-audio alignment strategy.}
 \label{tab:dataset}
\end{table*}

More specifically (see also Table~\ref{tab:dataset}), the reference is a CD recording made by Herbert von Karajan in 1985. To link this performance to the score, we manually annotated the recording at the bar level, involving 5,304 bar annotations (2,866 for the first, 2,438 for the second act).
The two tracking targets are recordings of live performances from the Vienna State Opera in 2018 and 2019, conducted by \'Adam~Fischer and Antonello~Manacorda, respectively, and performed by entirely different casts of singers. Again, for evaluation purposes, these had to be manually annotated, resulting in 10,608 new bar annotations.

For the study of this paper, we isolate from each recording the passages that contain only \textit{recitativo} sections. This subset represents between 16 and 18~\% of the total duration of each recording. In the experiments, the Fischer live recording will be used for the feature optimisation study; the Manacorda will then serve as the independent test set.

In total, the dataset comprises more than 9~hours of real-world opera music and involves a significant number of 15,912~manual bar-level annotations linking audio and pdf score. The score of ``Don Giovanni'' comprises some 530~pages, in the B\"arenreiter Urtext version (also available online, thanks to the \textit{Salzburg Mozarteum Foundation}\footnote{https://dme.mozarteum.at/DME/nma/}).

\begin{table}[t]
 \begin{center}
 \scalebox{1}{
 \begin{tabu} to \textwidth {@{} X[l] X[l] *4{X[c]}@{}}
  \toprule
  \textbf{Recording} & \textbf{dataset} & \textbf{mean} & $\mathbf{\leq 1s}$ & $\mathbf{\leq 2s}$ & $\mathbf{\leq 5s}$\ \\
  \midrule
  \textbf{Fischer} & full & 0.8s & 91.8\% & 95.0\% & 97.3\% \\
  \textbf{} & music & 0.2s & 97.2\% & 99.2\% & 99.9\% \\
  \textbf{} & recitative & 1.9s & 66.1\% & 75.3\% & 88.9\% \\
  \midrule
  \textbf{Manacorda} & full & 0.6s & 90.1\% & 94.5\% & 97.9\% \\
  \textbf{} & music & 0.2s & 96.2\% & 98.7\% & 99.7\% \\
  \textbf{} & recitative & 1.5s & 62.0\% & 77.2\% & 92.9\% \\
  \bottomrule
 \end{tabu}}
\end{center}
 \caption{SoTA tracker: Tracking error on full-length Don Giovanni, on the music subset only and on the recitative subset only. (Absolute durations in Fischer: full: 03:12:54; music: 02:37:56; recitative: 00:34:58.  Manacorda: full: 03:07:09; music: 02:36:29; Recitatives: 00:30:40.)}
 \label{tab:baseline_table}
\end{table}

\section{Baseline tracking evaluation}
\label{sec:baseline}

The opera tracker described in \cite{brazier2020towards} combines an On-Line Time Warping (OLTW) algorithm with audio event detectors (classifiers detecting the presence of applause, music and speech) that can halt the tracking process during transitions between parts. This makes the tracker quite robust to spontaneous events, such as applause, breaks and interludes, that frequently occur at transitions between arias or recitatives.
As input features to the DTW alignment algorithm, and based on a systematic study on classical orchestra recordings \cite{gadermaier2019study}, 120~MFCC features are extracted from both recordings; the first 20 are discarded, giving a feature vector of length 100 at each audio frame. FFT parameters are a window size of 20~ms and a hop size of 10~ms.

In order to highlight difficulties related to recitative tracking, we separately evaluate three different alignment scenarios: (1) alignment on the full-length target performances; (2) alignment on the music excluding from the opera all \textit{recitativo} sections (which leaves 4,665~annotated bars of music); and (3) alignment only on \textit{recitativo} sections (639~bars).

Table~\ref{tab:baseline_table} reports the tracking error on the Fischer and Manacorda performances, when aligned with the Karajan as the reference. The evaluation metrics used are standard in the field of score following \cite{cont2007evaluation}: for each alignment, we report the mean tracking error in seconds, over all bar annotations, as well as the cumulative proportion of bar positions tracked with an error of less than 1s, 2s, and 5s. We observe the same effect on both performances. For example, in the Fischer alignment, we have a mean tracking error of 0.8s, with 91.8\% of the bars detected with an error below 1 second, 95.0\% below 2s and 97.3\% below 5s. Tracking on the music-only part gives the lowest mean error and the highest percentages, implying a more precise alignment. The results on the recitative-only part are significantly worse: the number of bars detected with error below a second decreases to 66.1\% and 62.0\% for the Fischer and Manacorda, respectively.

Our goal in the following is to find a set of features that are better suited to accurate recitative alignment/tracking. In Section~\ref{sec:combination}, we will then investigate strategies for combining music- and recitative-focused features and trackers.

\section{Specific features for the Recitative Problem}
\label{sec:feature_study}

Our first step towards improving recitative tracking is to search for a set of features that are better suited to modeling passages with predominant voice content.
Operatic singing voices are usually clustered into six categories (bass, baritone, tenor, alto, mezzo-soprano, soprano), depending on their f0 (pitch) ranges. The frequency ranges extend from about 73~Hz to 1500~Hz \cite{villavicencio2015efficient, garnier2010vocal}. The features used in our baseline tracker and described in \cite{brazier2020towards} cover the frequency range [659~Hz;~22050~Hz]. Even if they enable accurate tracking in orchestra performances \cite{gadermaier2019study}, they may not be appropriate to track recitatives.

Aligning two performances that include the same linguistic content but whose vocal styles differ is not trivial. Apart from vocal effects such as vibratos or tremolos, a noticeable difference between speech and singing characteristics is the presence of a peak in the singing spectrum which is situated around 3~kHz, namely the 'singer's formant' \cite{sundberg1977acoustics}. Thus, and in order to get features robust to vocal style changes, one option is to restrict the frequency axis to a limit of 3~kHz to avoid this phenomenon and to focus on the frequency range mentioned previously.

\subsection{Feature Search Space}

In our search for an appropriate set of features, we again turn to Mel Frequency Cepstral Coefficients (MFCCs) and start a systematic search for a parametric setup optimized for recitative tracking. MFCCs are massively used in speech recognition and audio-to-lyrics alignment, and are already the features used in the baseline tracker. We extract MFCC features from three types of spectra that can be calculated in real time. The first spectrum is the standard one (called \textit{Spec} in the following), with a window size of 20~ms and a hop size of 10~ms. The two others are two different methods that estimate the \textit{spectral envelope} of the spectrum, the latter representing the phoneme information in a Source/Filter model (see equation (7) in \cite{gong2015real}). One method computes the Linear Predictive Coefficients (LPC), a robust feature that has already proven successful in the speech-to-singing alignment domain \cite{vijayan2018analysis}. The other computes the True Envelope (TE), which estimates the spectral envelope in an iterative way, and has been used in a singing-voice-to-score alignment task \cite{gong2015real}. From these two different spectral envelopes, we then extract the corresponding MFCCs.

To find the best set of parameters for the MFCC features, a grid search is conducted over the following parameter space:
\begin{itemize}
\item Sampling Rate: according to the previous considerations, we propose downsampling the initial audio signal. Six different \textbf{\#sr} values are tested: 1500, 3000, 6000, 12000, 24000 and 44100~Hz.
\item Number of MFCCs: this number changes the resolution of the features. Six \textbf{\#MFCC} values are proposed: 25, 50, 75, 100, 150, and 200.
\item Discarding Factor: this refers to the number of first (low-order) MFCCs to be discarded from the set. The factor was under study in \cite{gadermaier2019study}. Two \textbf{\#skip} values are proposed: 0 and 5.
\end{itemize}
The evaluation criterion for the grid search is the alignment error on the recitatives in the Fischer recording. The Manacorda is not used in this process, so that it can serve as an independent test set later.

In total, 216~experiments have been run, involving three types of features, six values of sampling rate, two discarding factors, and six numbers of MFCC.

\subsection{Optimisation on Fischer Recitatives}
\label{sec:testset_feature}

\begin{table}[t]
 \begin{center}
 \scalebox{1}{
 \begin{tabu} to \textwidth {@{} X[l] *5{X[c]}@{}}
  \toprule
  \textbf{feature} & \textbf{\#sr} & \textbf{\#MFCC} & \textbf{\#skip} & \textbf{mean (ms)} & $\mathbf{\leq 1s}$\\
  \midrule
  \textbf{baseline} & 44100 & 100 & 20 & 1915 & 66.1\%\\ \midrule
  \textbf{LPC} & 1500 & 25 & 0 & \textbf{955} & \textbf{76.5\%}\\
  \textbf{TE} & 1500 & 75 & 0 & 965 & 76.4\%\\
  \textbf{TE} & 1500 & 100 & 0 & 965 & 76.4\%\\
  \textbf{TE} & 1500 & 150 & 0 & 965 & 76.4\%\\
  \textbf{Spec} & 1500 & 75 & 0 & 966 & 76.4\%\\
  \textbf{TE} & 1500 & 200 & 0 & 976 & 76.2\%\\
  \textbf{Spec} & 1500 & 100 & 0 & 977 & 76.2\%\\
  \textbf{Spec} & 1500 & 150 & 0 & 977 & 76.2\%\\
  \textbf{Spec} & 1500 & 200 & 0 & 977 & 76.2\%\\
  \textbf{Spec} & 1500 & 50 & 0 & 979 & 76.2\%\\
  \textbf{Spec} & 1500 & 25 & 0 & 989 & 76.2\%\\
  \textbf{LPC} & 1500 & 50 & 0 & 958 & 76.1\%\\
  \bottomrule
 \end{tabu}}
\end{center}
 \caption{Feature parameter grid search on the Recitative subset of ``Don Giovanni'' conducted by \'Adam Fischer. Only the best results (percentage of bars detected with error $\le 1s$ higher than 76\%) are shown.}
 \label{tab:fischer_featuretable}
\end{table}

Table~\ref{tab:fischer_featuretable} shows the alignment errors on the Recitative subset of the Fischer performance (639~bars), for the 12 best feature parameter settings. As a reference, the \textit{baseline} features are the features used in our baseline tracker \cite{brazier2020towards}.

First, we note that all of these top 12 feature settings give relatively close results, in terms of mean alignment error and alignment precision. All features clearly outperform the baseline, reducing the mean error by one second and improving the percentage of precisely detected bars by at least 10 percentage points.

All the best results involve the same low sampling rate of 1500~Hz. This is in accordance with the argument made in the previous section. It allows not to include the 'singing formant' phenomenon in the different spectra and to zoom into the frequency range corresponding to the opera singers.

Finally, we note that increasing the discarding factor to 5 decreases significantly the alignment performance (not shown in the table), meaning that the lower order coefficients are important for recitative tracking. This relates to the point that the first MFCC coefficients are representative to the spectral shape \cite{majeed2015mel} which contain most of the phoneme information, according to the Source/Filter model.

A final observation (not included in Table \ref{tab:fischer_featuretable}) is that the \textit{maximum alignment error} over all these recitatives dropped from 26.8~seconds in the baseline case, to 13.1~seconds when using the optimised features. Thus, the new features may help reduce the risk of the tracker getting completely lost in difficult passages.

\subsection{Validating on Manacorda Recitatives}
\label{sec:validationset_feature}

Based on Table~\ref{tab:fischer_featuretable}, we conclude that the best performances were obtained in extracting a smoothed spectral envelope based on LPC with a sampling rate of 1500~Hz, 25~MFCC features, and a discarding factor of 0. We now validate this optimised feature set on the independent Manacorda recitative set and obtain the results shown in Table \ref{tab:manacorda_featuretable}. The improvement over the baseline is obvious. However, the amount of improvement is somewhat less spectacular than in the Fischer case; also the \textit{maximum} alignment error in this case remains unchanged, at 15.0 seconds, due to one particularly difficult situation in the recitative \textit{Ah ah ah ah, questa \`e buona}, which includes an improvised interlude of more than 13 seconds at the beginning of the third bar.

\begin{table}[t]
 \begin{center}
 \scalebox{1}{
 \begin{tabu} to \textwidth {@{} X[l] *5{X[c]}@{}}
  \toprule
  \textbf{feature} & \textbf{\#sr} & \textbf{\#MFCC} & \textbf{\#skip} & \textbf{mean (ms)} & $\mathbf{\leq 1s}$\\
  \midrule
  \textbf{baseline} & 44100 & 100 & 20 & 1503 & 62.0\%\\
  \textbf{LPC} & 1500 & 25 & 0 & \textbf{1023} & \textbf{69.7\%}\\
  \bottomrule
 \end{tabu}}
\end{center}
 \caption{Validating the features on the Recitative subset of the Manacorda performance.}
 \label{tab:manacorda_featuretable}
\end{table}

\section{Combining trackers}\label{sec:combination}

The goal of this work is to improve the accuracy of a full opera tracker. Based on the results from the previous section, we have now at our disposal two specific sets features: one that is effective for tracking orchestral music (the MFCCs from our baseline tracker), the other is effective for tracking recitatives. We now present and evaluate different strategies for combining both features in order to improve the overall opera tracking. To this end, we presuppose that we have two reasonably reliable (on-line) audio classifiers that detect the presence of music and speech, respectively, in an audio stream. Such classifiers, based on yet other sets of specialized features, are already components of our baseline tracker; they were trained on separate annotated training data and are briefly described in \cite{brazier2020towards}.

\begin{figure}[t]
\centering
\includegraphics[width=0.8\columnwidth]{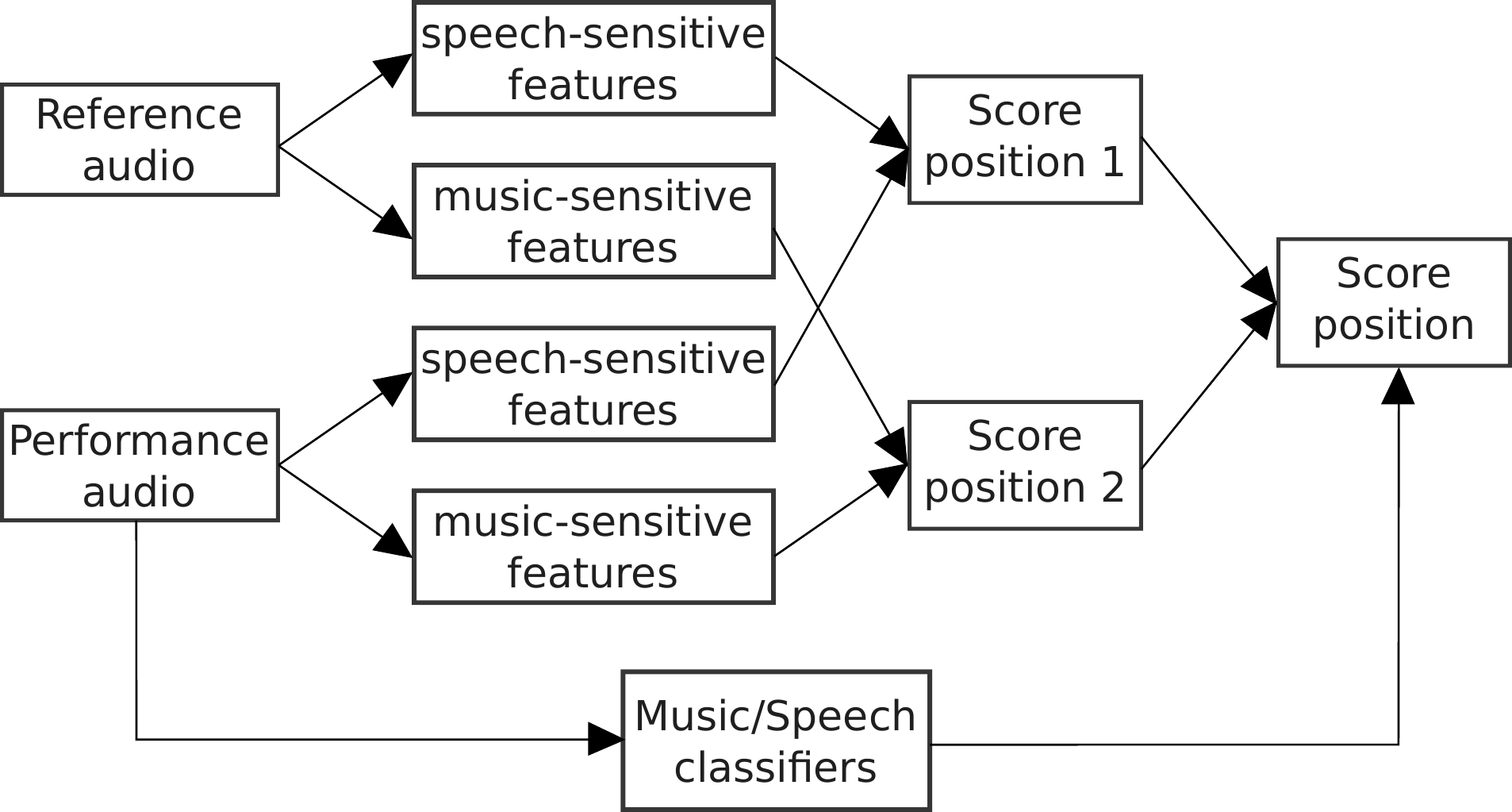}
\caption{Integrated Model.}
\label{fig:model}
\end{figure}

\subsection{Combination Strategies}

The combined tracking model is illustrated in Figure~\ref{fig:model}. From a performance, we extract a set of music-sensitive features and speech-sensitive features. From each type of feature, we compute the alignment between reference and performance, running two specialized OLTW trackers in parallel. At each audio frame, each DTW tracker outputs a cumulative distance vector whose minimal value is the hypothetical current score position. In the following we propose four simple solutions for combining these. We will use $\gamma_m$ to refer to the distance vector output by the music tracker, and $\gamma_s$ to the vector of the tracker using the speech-related (recitative) features.

\begin{enumerate}

\item \label{late_fusion}
Proposed by \cite{gong2015real}, the \textit{Late Fusion} strategy calculates the average of the two distance vectors $\gamma_{m}$ and $\gamma_{s}$. This strategy has proven to be successful in a Hidden semi-Markov Model. It has the advantage to compensate a weak tracker with the other one:

\begin{equation*} \label{eq:late_fusion}
\gamma_{\mathit{Late Fus}} = \frac{\gamma_{m} + \gamma_{s}}{2}
\end{equation*}

\item \label{speech_classifier}
The alternative combination variants we propose use our trained music/speech audio classifiers. In the \textit{speech-based fusion} strategy, considering that the tracker that aligns speech-sensitive features tends to be more accurate during the presence of speech, we weight the cumulative vectors by the probability of speech being present in the current signal, derived from the recent predictions of the speech detector:
\begin{equation*} \label{eq:speech_fusion}
\gamma_{\mathit{Speech Fus}} = (1 - w_{s}) \cdot \gamma_{m} + w_{s} \cdot \gamma_{s}
\end{equation*}

We calculate $w_{s}$ in two different ways: as the average of the last 25 speech predictions (probabilities) output by the speech detector (which, at a frame rate of 20 ms, gives a temporal context of 500ms), and as a weighted average with a linear weighting function from 0 to 1, with a context of 1s.

\item \label{music_classifier}
Analogously, the \textit{music-based fusion} strategy mixes in the music-sensitive tracker when music is detected:
\begin{equation*} \label{eq:music_fusion}
\gamma_{\mathit{Music Fus}} = w_{m} \cdot \gamma_{m} + (1 - w_{m}) \cdot \gamma_{s}
\end{equation*}

We also compute $w_{m}$ in two ways.

\item \label{music&speech_classifiers}
The \textit{music\&speech-based fusion} combines the two previous propositions and applies the equation:
\begin{eqnarray*} \label{eq:music&speech_fusion}
\gamma_{\mathit{M\&S Fus}} & = & 0.5 \cdot (w_{m}+(1-w_{s})) \cdot \gamma_{m} \\ & + & 0.5 \cdot ((1- w_{m})+w_{s}) \cdot \gamma_{s}
\end{eqnarray*}

\end{enumerate}

\subsection{Experiments}
\label{sec:strategy_experiments}

For the final experiments, we use the full-length Don Giovanni performance conducted by \'Adam Fischer to test the various combination strategies. In total, seven strategies are tested and evaluated on the 5,304 bar-level annotations. For each of these, we report the mean error and cumulative percentages for three error bounds.

\begin{table}[t]
 \begin{center}
 \scalebox{1}{
 \begin{tabu} to \textwidth {@{} X[l] *5{X[c]}@{}}
  \toprule
  \textbf{Strategy} & \textbf{mean type} & \textbf{mean} & $\mathbf{\leq 1s}$ & $\mathbf{\leq 2s}$ & $\mathbf{\leq 5s}$\ \\
  \midrule
  \textbf{baseline} & & 811ms & 91.8\% & 95.0\% & 97.3\% \\ \midrule
  \textbf{Late Fusion} &  & 384ms & 93.1\% & 96.6\% & 99.0\% \\ \midrule
  \textbf{Speech Fusion} & constant & 392ms & 92.9\% & 96.7\% & 98.9\% \\
  \textbf{} & weighted & 384ms & 93.0\% & 96.7\% & 99.0\% \\ \midrule
  \textbf{Music Fusion} & constant & 402ms & 92.9\% & 96.5\% & 98.9\% \\
  \textbf{} & weighted & 398ms & 92.9\% & 96.5\% & 98.9\% \\ \midrule
  \textbf{M\&S Fusion} & constant & \textbf{373ms} & \textbf{93.4\%} & \textbf{96.8\%} & \textbf{99.0\%} \\
  \textbf{} & weighted & 374ms & 93.3\% & \textbf{96.8\%} & 99.0\% \\
  \bottomrule
 \end{tabu}}
\end{center}
 \caption{Tracking errors on full-length Don Giovanni conducted by \'A.~Fischer, with different strategies to combine the two parallel trackers.}
 \label{tab:fischer_strategytable}
\end{table}

Table~\ref{tab:fischer_strategytable} shows the results. Clearly, all combinations outperform the baseline, with the mean error being more than halved. The type of mean used to smooth the music/speech detector predictions does not have much influence. However, the combined music\&speech-based fusion strategy outperforms all of the other methods. The best results are obtained with a constant average for the weights, with a mean error of 373ms and an increase of the different percentages by more than 1.6 percentage points compared to the baseline. We thus select this strategy in the following.

As a final test, we apply the combined tracker to our independent test piece -- the Manacorda performance (using the Karajan as the score reference, as always), and hope to see similar improvement as in Table~\ref{tab:fischer_strategytable}.

\begin{table}[t]
 \begin{center}
 \scalebox{1}{
 \begin{tabu} to \textwidth {@{} X[l] *4{X[c]}@{}}
  \toprule
  \textbf{Strategy} & \textbf{mean} & $\mathbf{\leq 1s}$ & $\mathbf{\leq 2s}$ & $\mathbf{\leq 5s}$\ \\
  \midrule
  \textbf{baseline} & 561ms & 90.1\% & 94.5\% & 97.9\% \\
  \textbf{M\&S Fusion} & \textbf{547ms} & \textbf{90.3\%} & \textbf{94.7\%} & \textbf{98.0\%}\\
  \bottomrule
 \end{tabu}}
\end{center}
 \caption{Error on full-length Don Giovanni conducted by Manacorda, using the combination of two parallel trackers.}
 \label{tab:manacorda_strategytable}
\end{table}

The result shown in Table~\ref{tab:manacorda_strategytable} is somewhat sobering: the improvement brought about by the Music\&Speech strategy is minimal. As the alignment results on the recitative-only part did show a substantial improvement on the Manacorda (see Table~\ref{tab:manacorda_featuretable}), we conclude that it is not the recitative-specific features that prevent an overall improvement, but that it must be the combination strategy, or the audio classifiers (music/speech) controlling it. Recitatives are sonically extremely diverse in different recordings, especially regarding singing and speaking style. One immediate goal of future research will be to perform broader and more detailed investigations into classifiers that can reliably distinguish aria-style singing from recitative-style singing, which will, first of all, again require curating a broader set of annotated opera data.

\section{Discussion \& Conclusion}\label{sec:conclusion}

This paper has focused on the problems encountered by a state-of-the-art opera tracking algorithm when faced with \textit{recitativo} sections. We isolated these sections in two live recordings of a full-length typical opera, and tuned specific features in order to improve the accuracy of the alignment. Then we proposed a new method for opera tracking that involves two trackers working in parallel, one relying on speech-sensitive features, the other on music-sensitive features, and a way of combining these based on on-line music and speech classifiers. We validated the proposed method on a complete unseen opera, with (mostly) encouraging results. However, there remain a number of open problems.

The results reported in Table~\ref{tab:fischer_featuretable} and Table~\ref{tab:manacorda_featuretable} show that the optimized features help to increase by 8 and 10\% the number of bars detected with an error less than one second in the Recitative corpus, to a level of about 70 \%. This is still not at the same level of precision at which we can track non-recitative passages (cf.~Table~\ref{tab:baseline_table}), and may not be sufficient for all kinds of desirable applications (such as automated subtitle display in the opera hall, or live video cutting for live streaming). In describing the numerous challenges arising in recitative tracking, we emphasized that the only reliable invariant among different performances is the linguistic content. Thus, audio-to-lyrics alignment would be a promising direction to pursue, but this implies new challenges. There is no freely available opera corpus for training appropriate phoneme recognition models; such training corpora would likely have to be singing- and language-specific. For instance, available Italian corpora, such as MLSPKA, CommonVoice, or Damp Dataset involve only speech signals and karaoke songs. Thus, we will again have to curate our own training data; as a next step, we plan to annotate full-length operas at the sentence level in order to train such audio-to-lyrics models.

Another problem, which was hinted at in Section~\ref{sec:data}, concerns \textit{structural mismatches} (e.g., parts omitted or variants inserted) between score (reference) and performance (target). As a future task, we plan to address this problem by adapting existing methods (e.g., \textit{JumpDTW} \cite{fremerey2010handling}) to a real-time scenario.

The present study was conducted on only one Opera\footnote{We are aware of the fact that a single opera (in three different recordings) may seem like rather little data. However, it should be kept in mind that a full opera represents several hours of music and requires massive annotation work. For instance, the score of `Don Giovanni' is a book of 500 pages, containing 5304 bars of music.}, Don Giovanni, in three different versions. Over time, more operas from other composers, periods, and genres will need to be investigated. Annotating even one full opera implies a lot of work, in our case involving more than 15,000~bar-level annotations over three performances, and it would have been nice to provide the resulting dataset to the scientific community. Unfortunately, the rights to our Don Giovanni recordings are with the Vienna State Opera. However, we have recently been involved in an amateur recording of Mozart's complete \textit{Zauberfl\"ote}, which we are in the process of annotating (including lyrics), and which will be made available to the community in the near future.

\begin{acknowledgement}
This project has received funding from the European Union’s Horizon 2020 research and innovation programme under the Marie Sklodowsa-Curie grant agreement No. 765068 (Project MIP-Frontiers). Special thanks to Christopher Widauer for providing the Don Giovanni recordings from the Vienna State Opera.
\end{acknowledgement}
%

\bibliographystyle{plain}
\bibliography{references}

\end{document}